\documentclass{article}

\usepackage{amsfonts}
\usepackage{amsmath}
\usepackage{amssymb}
\usepackage{amsthm}
\usepackage{enumerate}

\newtheorem{theorem}{Theorem}[section]

\theoremstyle{definition}

\newtheorem{remark}[theorem]{Remark}
\newtheorem{example}{Example}

\newcommand{\vZ}{\mathbb{Z}}
\newcommand{\vF}{\mathbb{F}}

\newcommand{\user}{\mathcal{U}}
\newcommand{\attacker}{\mathcal{M}}

\begin{document}

\title{An Active Attack on a Multiparty Key Exchange Protocol%
\thanks{The Research was supported in part by the Swiss National Science
Foundation under grant No.\@ 149716. First author is partially
suppported by Ministerio de Educacion, Cultura y Deporte grant
``Salvador de Madariaga'' PRX14/00121, Ministerio de Economia y
Competitividad grant MTM2014-54439 and Junta de Andalucia (FQM0211).
The first author is supported by Armasuisse.
}
}

\author{R. Schnyder\footnote{University of Zurich}, J.A. L\'opez-Ramos\footnote{University of Almeria}, J. Rosenthal\footnotemark[2], D. Schipani\footnotemark[2]}

\maketitle
\begin{abstract}
The multiparty key exchange introduced in Steiner et al.\@ and
presented in more general form by the authors is known to be secure
against passive attacks.
In this paper, an active attack is presented assuming malicious
control of the communications of the last two users for the duration
of only the key exchange.
\end{abstract}

\section{Introduction}

The increased use of light and mobile devices has led to the study of
the so called mobile ad hoc networks.
These are created, operated and managed by the nodes themselves and
therefore are solely dependent upon the cooperative and trusting
nature of the nodes.
The ad hoc property of these mobile networks implies that the network
is formed in an unplanned manner to meet an immediate demand and
specific goal, and that the nodes are continuously joining or leaving the
network.
Thus, key management in this type of networks is a very important
issue and has been the aim of numerous works since then
(see~\cite{lee} or~\cite{vandenmerwe} and their references).

One of the most widely known such schemes is due to Steiner et
al.\@ and is known as Cliques (cf.~\cite{steiner2}).
Cliques is a multiparty key exchange protocol generalizing the
Diffie-Hellman key exchange based on the discrete logarithm problem.
It is composed of an initial key agreement (IKA) to set up a first
common key and an auxiliary key agreement (AKA) in order to refresh
the key at any later stage.

In~\cite{pereira}, the authors propose a systematic way for analyzing
protocol suites which extend the Diffie-Hellman key-exchange scheme
to a group setting.
They find interesting attacks which exploit algebraic properties of
Diffie-Hellman exponentiation.
However, our attack uses a different approach that exploits a weakness
of a specific protocol and allows for prolonged eavesdropping.

We will consider in particular one of the proposed initial key
agreements referred to (in~\cite{steiner2}) as IKA.2.
The authors generalize these schemes in~\cite{lopez}, considering a
general action on a semigroup, and this is how IKA.2 is presented
below.

We will then show an active attack on this protocol that requires
control of the communications of two particular parties for only the
duration of the key exchange.
That is, unlike in a regular man-in-the-middle attack, it is not
necessary for the attacker to control the communications after the key
exchange in order to translate messages, since all users are made to
agree on the same key.

Although it is not possible for the attacker to keep a copy of the key
after the users initiate AKA operations, we will show how she can
avoid being noticed at that point.

\section{An Initial Key Agreement protocol}
\label{sec:protocol}

The protocol below gives $n$ users the possibility to share an initial
common key built using their private keys.
A proof of its correctness and security against passive attacks can be
found in~\cite{lopez,steiner2}, assuming the Diffie-Hellman problem is
hard for the given group action.

Suppose we have $n$ users $\user_1, \ldots, \user_n$ who wish to agree
upon a common key.

Let $G$ be an abelian group, written multiplicatively.
Let $S$ be a set, and suppose we have a group action
\begin{align*}
	G \times S &\to S \\
	(g, s) &\mapsto g \cdot s.
\end{align*}

The users publicly agree on a common element $C_0 =s \in S$, and for
each $i = 1, \ldots, n$, the user $\user_i$ selects a secret group
element $g_i \in G$.

The protocol proceeds as follows:
\begin{enumerate}[(1)]
	\item For $i = 1, \ldots, n-2$,\; $\user_i$ sends to $\user_{i+1}$
		the message $C_i = g_i \cdot C_{i-1}$.
	\item $\user_{n-1}$ broadcasts $C_{n-1} = g_{n-1} \cdot C_{n-2}$ to
		the other users $\user_1 , \ldots ,\user_{n-2}, \user_n$.
	\item $\user_n$ computes the shared key $K = g_n \cdot C_{n-1}$.
	\item For $i = 1, \ldots, n-1$,\; $\user_i$ sends $D_i = g_i^{-1}
		\cdot C_{n-1}$ to $\user_n$.
	\item $\user_n$ broadcasts $\{ g_n \cdot D_1, g_n \cdot D_2, \ldots,
		g_n \cdot D_{n-1}, C_{n-1} \}$ to $\user_i$, $i=1, \ldots, n-1$.
	\item For $i = 1, \ldots, n-1$,\; $\user_i$ computes the shared key
		$K = g_i \cdot (g_n \cdot D_i)$.
\end{enumerate}

It is easy to see that for $i = 1, \ldots, n-1$, we have that
\begin{align*}
	C_i &= \bigg( \prod_{j=1}^i g_j \bigg) \cdot s, \\
	D_i &= \bigg( \prod_{\substack{j=1\\j\ne i}}^{n-1} g_j \bigg) \cdot
		s,
\intertext{and finally}
	K = C_n &= \bigg( \prod_{j=1}^n g_j \bigg) \cdot s.
\end{align*}

From the above, we can also observe that $C_{n-1}$ is not needed by
any user to recover the session key $K$.
However, this information is disclosed for future rekeying purposes,
as we will see later.

\begin{example}
	Let $\vF_q$ be a finite field.
	Let us consider an element $g$ of prime order $p$, generating the
	subgroup $S \subset \vF_q^*$.
	Then the action $\Phi\colon \vZ_p^* \times S \to S$ defined by
	$\Phi(x, h) = h^x$ provides the Initial Key Agreement protocol
	introduced in~\cite[Section~4.2]{steiner2} as IKA.2.
\end{example}

\begin{example}
	Let us denote by $\varepsilon$ the group of points of an elliptic
	curve of prime order $p$.
	Then the action $\Phi\colon \vZ_p^* \times \varepsilon \to
	\varepsilon$ defined by $\Phi(x,P) = xP$ gives an elliptic curve
	version of IKA.2 cited above.
\end{example}

\section{An active attack on the Initial Key Agreement}
\label{sec:active_attack}

We describe an active attack on the protocol of the preceding section.
Suppose that the attacker $\attacker$ wants the users $\user_1,
\ldots, \user_n$ to agree on a shared key as usual, except that she
is in possession of the key as well.

In order to carry out our attack, $\attacker$ needs to have full
control over the communication of the users $\user_{n-1}$ and
$\user_n$ for the duration of the key exchange.
However, unlike in a regular man-in-the-middle attack, she does not
need to maintain this control after the key exchange is completed.

In the beginning, $\attacker$ chooses her own secret group element
$\hat{g} \in G$.
She then proceeds as follows:

\begin{enumerate}[(a)]
	\item Step (1) is carried out as usual.
	\item $\attacker$ intercepts the broadcast of $\user_{n-1}$ during
		step (2) and remembers the value $C_{n-1}$.
		At this point, all users except for $\user_{n-1}$ are sitting in
		step (2), waiting for the broadcast that was halted.
	\item $\user_{n-1}$ proceeds to step (4), where he sends
		$g_{n-1}^{-1} \cdot C_{n-1} = C_{n-2}$ to $\user_n$.
		This is also intercepted by $\attacker$.
		$\user_{n-1}$ is now waiting in step (5).
	\item $\attacker$ now makes $\user_n$ believe that he received the
		broadcast of step (2), but actually sends him $\hat{g} \cdot
		C_{n-1}$.
		At this point, $\user_n$ computes the shared key $K = g_n \hat{g}
		\cdot C_{n-1}$ and waits in step (4).
	\item $\attacker$ now sends to $\user_n$ the values $\{ m_1, \ldots,
		m_{n-3}, C_{n-2}, C_{n-1} \}$, pretending that they were sent by
		the other users in step (4).
		The $m_i$ are random elements of the orbit $G \cdot s$.
	\item In step (5), $\user_n$ sends back, among others, the values
		$g_n \cdot C_{n-2}$ and $g_n \cdot C_{n-1}$, which $\attacker$
		intercepts.
		The user $\user_n$ is now finished, and $\attacker$ can compute
		the shared key $K = \hat{g} g_n \cdot C_{n-1}$.
	\item Until now, $\user_1, \ldots, \user_{n-2}$ have been waiting
		for the broadcast in step (2), which $\attacker$ now provides in
		the form of $g_n \cdot C_{n-1}$.
	\item $\user_i$, $i = 1 \ldots, n-2$, go to step (4) and send back
		$g_i^{-1} g_n \cdot C_{n-1}$, which $\attacker$ intercepts.
	\item In step (5), $\attacker$ broadcasts to $\user_i$, $i = 1
		\ldots, n-2$, the message
		\begin{equation*}
			\{ \hat{g} g_1^{-1} g_n \cdot C_{n-1}, \hat{g} g_2^{-1} g_n
			\cdot C_{n-1}, \ldots, \hat{g} g_{n-1}^{-1} g_n \cdot C_{n-1},
			g_n \cdot C_{n-1} \}
		\end{equation*}
		User $\user_{n-1}$ is sent the same message, but
		the last element, $g_n \cdot C_{n-1}$ is substituted by $C_{n-1}$.
	\item The users $\user_1, \ldots, \user_{n-2}$ now all compute the
		shared secret $K = g_i \hat{g} g_i^{-1} g_n \cdot C_{n-1}$.
\end{enumerate}

Let us make some comments on the attack introduced above.
First, we can observe that at the end of this procedure, all users as
well as the attacker share the same key
\begin{equation*}
	K = \bigg( \prod_{j=1}^n g_j \bigg) \cdot ( \hat{g} \cdot s).
\end{equation*}
Any passive observer will still be unable to determine the key, for
the same reason that the original protocol is secure against passive
attacks, cf.~\cite[Theorem~2.1]{steiner2}, whose proof also applies
to the general setting given in Section~\ref{sec:protocol} whenever
the action is transitive and the Diffie-Hellman problem is hard.

The attacker's secret $\hat{g}$ is not strictly required for the
attack to work, but without it, the users may notice that something is
amiss.
Namely, in step (e), if we leave out $\hat{g}$, the user $\user_n$ may
notice that $\attacker$ sent the same value $C_{n-1}$ as in step (d).
Similarly, in step (i), the other users could notice that the attacker
just returned their transmission from (h).
Using $\hat{g}$, however, the users should be unable to tell the
difference between a regular execution of the protocol and the attack,
again as a consequence of~\cite[Theorem~2.1]{steiner2}.

As in the Initial Key Agreement (IKA) protocol introduced in
Section~\ref{sec:protocol}, the broadcast element $g_n \cdot
C_{n-1}$ is added at the end of the message in (i) in view of future
rekeying operations and is not needed by any of the users $\user_1,
\ldots, \user_{n-2}$ to recover the shared key.
Note that users $\user_i$, $i=1, \ldots, n-2$, expect that the last
element of the message sent in step (i) is the one broadcast in step
(2) of the protocol, which the attacker substitutes precisely by $g_n
\cdot C_{n-1}$.
In the case of user $\user_{n-1}$, who is also expecting the element
sent in step (2) of the protocol, the element that $\attacker$ sends
in step (b) is $C_{n-1}$.
If this is not satisfied, the users might notice that something is
wrong.

\section{An exit strategy}
\label{sec:exit_strategy}

After the attack of Section~\ref{sec:active_attack}, the attacker
$\attacker$ shares the key with the users $\user_1, \ldots, \user_n$
and can listen in on their conversation without any further active
measures.
However, at some point after that, the users may wish to execute an
AKA operation, which is to say a key refreshment, the addition of a
new member to the group, etc.\@ as described
in~\cite[Section~5]{steiner2}.
After this point, the attacker can certainly no longer listen to the
conversation.
Even worse, the values the users remember from step (5) of the
protocol are substantially different from normal, and any key refresh
operation will thus fail completely, alerting the users about the
attack.

In what follows, we will describe how the attacker can avoid being
noticed by forging key refresh operations herself, assuming that any
user may initiate a key refreshment at any time.

First, we recall the key refresh operation after a regular execution
of IKA.2, adapted from~\cite[Section~5.6]{steiner2}.
Suppose user $\user_c$ wishes to initiate a key refreshment.
He remembers from step (5) of the key agreement protocol the values
$\{ E_1, \ldots, E_n \}$, where $E_k = \big( \prod_{j=1, j\ne k}^{n}
g_j \big) \cdot s$, $k = 1, \ldots, n$.
He picks a new secret $g_c' \in G$ and broadcasts
\begin{equation*}
	\{ g_c' \cdot E_1, \ldots, g_c' \cdot E_{c-1}, E_c, g_c' \cdot
	E_{c+1}, \ldots, g_c' \cdot E_n \}.
\end{equation*}
Now, all users can compute the new key $g_c' \cdot C_n = g_c' \cdot
\big( \prod_{j=1}^n g_j \big) \cdot s$.
User $\user_c$ also replaces his own secret with $g_c' g_c$, and
everyone replaces the information remembered from step (5) with this
new broadcast.

\begin{remark}
	One important detail to note is that when $\user_c$ initiates the
	key refreshment, the value $E_c$ he sends in position $c$ is
	unchanged and already known to the other users.
	Hence, if $\attacker$ wishes to forge a key refreshment coming from
	$\user_c$, she has to make sure that each user receives in position
	$c$ the value he previously held there.
	Otherwise, the attack could be discovered.
\end{remark}

Suppose now that the attacker $\attacker$ has just executed the attack
from Section~\ref{sec:active_attack}.
Instead of $\{ E_1, \ldots, E_n \}$, the users now remember the
following values:
\begin{itemize}
\item For $i=1, \ldots, n-2$, $\user_i$ remembers $\{ \hat{g} \cdot
	E_1, \ldots, \hat{g} \cdot E_{n-1}, C_n \}$.
\item $\user_{n-1}$ remembers $\{ \hat{g} \cdot E_1, \ldots , \hat{g}
	\cdot E_{n-1}, E_n \}$.
\item $\user_n$ remembers $\{ g_n \cdot m_1, \ldots, g_n \cdot
	m_{n-3}, E_{n-1}, C_n, \hat{g} \cdot E_n \}$.
\end{itemize}

Evidently, if some user tries to initiate a key refreshment with these
values, the operation will fail.
However, $\attacker$ can bring the users into a consistent state by
forging two key refresh operations herself.
For this, she needs to still have control over the communications of
$\user_{n-1}$ and $\user_n$, as in the original attack.

First, $\attacker$ picks two new random values $\hat{f}$ and $\hat{h}
\in G$.
Then, she forges a key refresh operation by sending the following
values to the different users:
\begin{itemize}
	\item To $\user_i$, $i = 1, \ldots, n-2$, she sends
		\begin{equation*}
			\{ \hat{h} \hat{g} \cdot E_1, \hat{h} \hat{g} \cdot E_2, \ldots,
			\hat{h} \hat{g} \cdot E_{n-2}, \hat{g} \cdot E_{n-1}, \hat{f}
			\hat{h} \hat{g} \cdot E_n \},
		\end{equation*}
		pretending it came from $\user_{n-1}$.
	\item To $\user_{n-1}$, she sends
		\begin{equation*}
			\{ \hat{f} \hat{h} \hat{g} \cdot E_1, \hat{h} \hat{g} \cdot E_2,
			\ldots, \hat{h} \hat{g} \cdot E_{n-2}, \hat{h} \hat{g} \cdot
			E_{n-1}, E_n \},
		\end{equation*}
		pretending it came from $\user_n$.
	\item To $\user_n$, she sends
		\begin{equation*}
			\{ \hat{f} \hat{h} \hat{g} \cdot E_1, \hat{h} \hat{g} \cdot E_2,
			\ldots, \hat{h} \hat{g} \cdot E_{n-2}, C_n, \hat{h} \hat{g} E_n
			\},
		\end{equation*}
		pretending it came from $\user_{n-1}$.
\end{itemize}
After this, the users will agree on the shared key $\hat{h} \hat{g}
\cdot C_n$, which is also known to $\attacker$.
As remarked above, if a user is made to believe that he received a key
refreshment from $\user_c$, he must receive in position $c$ the value
he already held there.

Now, the values held by the users are still inconsistent, so
$\attacker$ has to forge a second key refreshment:
\begin{itemize}
	\item To $\user_i$, $i = 1, \ldots, n-2$, she sends
		\begin{equation*}
			\{ \hat{f} \hat{h} \hat{g} \cdot E_1, \ldots,
			\hat{f} \hat{h} \hat{g} \cdot E_n \},
		\end{equation*}
		pretending it came from $\user_n$.
	\item To $\user_{n-1}$ and $\user_n$, she sends
		\begin{equation*}
			\{ \hat{f} \hat{h} \hat{g} \cdot E_1, \ldots,
			\hat{f} \hat{h} \hat{g} \cdot E_n \},
		\end{equation*}
		pretending it came from $\user_1$.
\end{itemize}
Now, all users and the attacker agree on the shared key $\hat{f}
\hat{h} \hat{g} \cdot C_n$.
Furthermore, all users remember the same consistent values for key
refreshment.
If in the future any user initiates a key refreshment or other AKA
operation, the attacker will lose access to the key, but the operation
itself will work out without problem and without the users noticing
anything wrong.

\begin{remark}\label{rem:mitm}
	An alternative course of action for $\attacker$ is to convert the
	attack into a regular man-in-the-middle attack on $\user_n$ at the
	time of the first key refreshment.
	For this, note that given the values each user remembers, a key
	refreshment initiated by $\user_c$, $c \le n-2$, works well for all
	users but $\user_n$.
	The attacker can then intercept the broadcast arriving at $\user_n$
	and replace it with random values, except that at position $n$ she
	sends $\hat{h} \cdot E_n$ for some random $\hat{h} \in G$, and at
	position $c$ she sends $g_n \cdot m_c$, which she knows from step
	(f) of the attack.
	Then, $\attacker$ will have the key $\hat{g} g_c' \cdot C_n$ in
	common with $\user_i$, $i \le n-1$, as well as $\hat{h} \cdot C_n$
	with $\user_n$.
	From then on, she can run a regular man-in-the-middle attack.
	A similar attack can be carried out if $\user_n$ initiates a key
	refreshment, but not if $\user_{n-1}$ does so.
	In this case, the attacker can intercept and apply $\hat{g}$ to the
	message for $\user_{n}$ so that all users agree on a common key
	without noticing the previous attack.
\end{remark}

\end{document}